\documentclass[aps,prd,twocolumn,superscriptaddress,nofootinbib,showpacs,preprintnumbers]{revtex4}
\usepackage[english]{babel}
\usepackage{graphicx}

\begin{document}

\title{A dark energy multiverse}
\author{Salvador Robles-P\'erez}
\email{salvarp@imaff.cfmac.csic.es}

\author{Prado Mart\'{\i}n-Moruno}
\email{pra@imaff.cfmac.csic.es}

\author{Alberto Rozas-Fern\'andez}
\email{a.rozas@imaff.cfmac.csic.es}

\author{Pedro F. Gonz\'alez-D\'\i az}
\email{p.gonzalezdiaz@imaff.cfmac.csic.es}

\affiliation{Colina de
los Chopos, Instituto de Matem\'aticas y F\'\i
sica Fundamental, \\
Consejo Superior de Investigaciones Cient\'\i ficas, Serrano 121,
28006 Madrid, Spain}
\date{\today}

\begin{abstract}
We present cosmic solutions corresponding to universes filled with dark and phantom energy, all having a negative cosmological constant. All such solutions contain infinite singularities, successively and equally distributed along time, which can be either big bang/crunchs or big rips singularities. Classicaly these solutions can be regarded as associated with multiverse scenarios, being those corresponding to phantom energy that may describe the current accelerating universe.
\end{abstract}

\preprint{IMAFF-RCA-07-01}
\maketitle

Just like the word atom designated what in principle was thought
to be indivisible and finally turned out not to be the case; the
word universe, which was originally intended to describe the
whole, has recently been reinterpreted to be just a single
causally disconnected part from the whole spacetime. Different
spacetimes could exist and our universe would be just one more
among that ensemble of completely causally disconnected
spacetimes. Actually, it was Giordano Bruno who first realized that there could exist many other worlds other than ours \cite{Giordano}. This idea has triggered several centuries later the development of different
theories of the multiverse, this time with quite less risks. Quite possibly, the best known is the
many-universes theory derived from the relative-state-formulation
due to Everett when applied to cosmology
\cite{Everett:1957,DeWitt}, which states that all branches of a
wave function for the universe correspond to equally real
different universes existing in parallel within an overall
multiverse.

But there are multiverse models, too that appear outside the
quantum realm, in the framework of general relativity. One example
of a multiverse that does not make explicit recourse to a quantum
formalism could be the chaotic inflationary multiverse \cite{Linde}. In every
flat space which has an event horizon, such as it happens in the
inflationary universe, a closed causal region of spacetime is
settled which can be influenced by observers. Since the universe
is flat, it is infinite so for observers who are space-like
separated by distances greater than the sum of their respective
distances to the event horizons, their respective causal domains
are disjoint and therefore every inflationary domain can be
interpreted as a single universe in the framework of this
classical multiverse. Another possible multiverse may appear when
we consider the current accelerated expansion of the Universe. If
we choose as dark energy phantom energy \cite{Caldwell:1999ew},
then a singularity is predicted to occur in the finite future
\cite{Caldwell:2003vq}. This singularity divides the universe into
two classically non-connected regions, before and after the
singularity \cite{Salva}. Here an idea of the multiverse would
also appear because that model necessarily requires a precise
discretization of the parameter in the equation of state, if one
wants to consider the region after the big rip as a part of the
whole spacetime. Each value of the discretized parameter of the
equation of state would then describe a single universe in the
context of an in infinite multiverse.

Recently, string theory has also resorted to the multiverse idea
to interpret the multiplicity of positive- energy vacua which
rises up to order $10^{100}$ to $10^{500}$ \cite{Ashok}. The different
subuniverses described by this string landscape \cite{Susskind} could be different
regions of space, different eras of time in a single big bang,
different regions of spacetime or different parts of quantum
mechanical Hilbert space ( being these alternatives not mutually
exclusive) \cite{Weinberg:2005fh}.

Furthermore, another multiverse model has been discussed by
Smolin \cite{Smolin}, who conjectured that new universes are spawned within
black holes, and that this kind of baby universes will inherit the
physics of the parent universe but with small random variations.
The process could continue ad infinitum. Universes that produce
many black holes would induce more progeny too, representing the
largest volume of space.

On the other hand, in the ekpyrotic model of Steinhard and Turok \cite{Steinhard},
a brane collides with a confining three-dimensional boundary to a
four-dimensional space to create the big bang. The
four-dimensional space can be foliated with any number of branes
each of which, in the absence of collisions, constitutes a
universe.

Within the framework of the current accelerated expansion of the
universe mentioned above, we have considered in this paper a new
model in which we have taken into account the existence of a
negative cosmological constant. A spacetime with a negative
cosmological constant is worth investigating, since it allows a
consistent physical interpretation and naturally appears in
elementary particle theories. Indeed, in string theory, as in
supergravity theories, the vacuum has a negative energy density,
which means that it is described by anti-de Sitter (AdS)
spacetime. In an important advance to understand quantum issues in
strong gravitational fields it was conjectured in 1997 that string
theory in an AdS background is equivalent to a conformal field
theory (CFT) \cite{Maldacena}. This is a beautiful and concrete example of the
holographic principle in quantum gravity \cite{thooft}. It is with this
motivation that we in this paper consider a cosmic model of dark
energy with a negative constant vacuum energy.

If we consider a quintessence field to describe dark energy, one
can describe it as a perfect fluid with an equation of state
$p=w\rho=w\rho_0(a(t)/a_0)^{-3(1+w)}$, where $p$ and $\rho$ are
the pressure and energy density of the fluid, respectively, and
$w$ a constant parameter. The Friedmann equation for this flat
model, which contains a negative cosmological constant $\Lambda$,
can be written as
\begin{equation}\label{uno}
H^2=-\lambda+Ca^{-3\beta},
\end{equation}
with $\lambda=|\Lambda|/3$, where $\lambda<8\pi\rho_0/3$ in order
for $H_0$ to be real; $C=8\pi\rho_0/(3a_0^{-3\beta})$ and
$\beta=1+w$. By integrating Eq. (\ref{uno}), we can obtain the
cosmic scale factor, yielding
\begin{eqnarray}\label{dos}
a(t)&=&a_0\left[\cos\left(\frac{3\beta}{2}\lambda^{1/2}(t-t_0)\right)\right.\nonumber\\&&\left.+\left(\frac{C}{\lambda}a_0^{-3\beta}
-1\right)^{1/2}\sin\left(\frac{3\beta}{2}\lambda^{1/2}(t-t_0)\right)\right]^{\frac{2}{3\beta}}.
\end{eqnarray}

For the case in which the dark energy is phantom energy, that is,
when $\beta<0$, this factor is converted into
\begin{equation}\label{tres}
a(t)=a_0\left[\cos\left(\alpha(t-t_0)\right)-b\sin\left(\alpha(t-t_0)\right)\right]^{-\frac{2}{3|\beta|}},
\end{equation}
where $\alpha=\frac{3|\beta|}{2}\lambda^{1/2}$ and
$b=\left(\frac{8\pi}{3\lambda}\rho_0-1\right)^{1/2}$. It is easy
to see that the scale factor diverges an infinite number of times
along the full time interval. Each of such divergences actually
describes a big rip singularity that takes place at
\begin{equation}\label{cuatro}
t_{br_{n}}=t_0+\frac{2}{3|\beta|\lambda^{1/2}}{\rm
arctg}\left[\left(\frac{8\pi\rho_0}{3\lambda}-1\right)^{-1/2}\right]+\frac{2n\pi}{3|\beta|\lambda^{1/2}},
\end{equation}
with $n$ any natural number. We recover the expression for the big
rip time obtained in a quintessence model of phantom energy
without cosmological constant \cite{Caldwell:2003vq}, when we set
$n=0$ in expression (\ref{cuatro}) and expand it for $\lambda<<1$,
\begin{equation}\label{cinco}
t_{br}=t_0+\frac{1}{|\beta|\left(6\pi\rho_0\right)^{1/2}}.
\end{equation}

\begin{figure}[h]
\begin{flushleft}
\includegraphics[width=0.8\columnwidth]{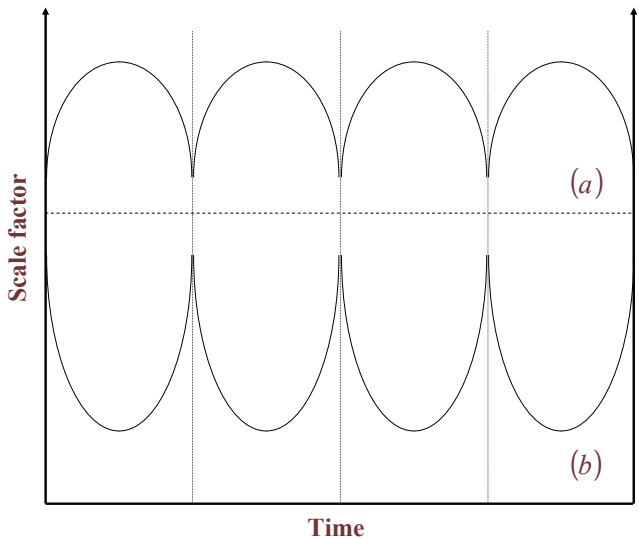}
\end{flushleft}
\caption{Time evolution corresponding to a universe equipped with a negative cosmological constant and: $(a)$ quintessential dark energy with $\beta>0$ and $(b)$ phantom energy with $\beta<0$.}
\label{univ}
\end{figure}

In the light of Eq. (\ref{cuatro}) we can in fact see that this
model will have infinite big rip singularities. This can be
interpreted as follows: classically, a singularity cuts off the
space time, so the different regions between big rips would be
isolated. Thus, each of them would correspond to a different
universe, independent of the rest, i.e., another spacetime. But,
as Eq.(\ref{tres}) tell us, these independent universes are
identical among them and have the same physical characteristics.
All of them begin at a big rip singularity, and then progressively
contract until a given, constant, minimum value of the scale
factor,
\begin{equation}\label{seis}
a_{{\rm min}}=a_0
\left(\frac{8\pi\rho_0}{3\lambda}\right)^{1/3\beta}>0,
\end{equation}
after which the given universe starts expanding, all the way in an
accelerated fashion, to again reach the next big rip singularity,
(see Fig.~\ref{univ}). The minimum value in Eq. (\ref{seis}) has
been obtained from the extremum value that corresponds to equating
to zero Eq. (\ref{uno}). The lifetime of every of these universes
is given by
\begin{equation}\label{siete}
t_u=\frac{2\pi}{3|\beta|\lambda^{1/2}}.
\end{equation}
It follows from Eq. (\ref{siete}) that the smaller $\lambda$ the
longer the universe life $t_u$. It can be seen that if
$\lambda=0$, where we recover the quintessence model of phantom
energy, these time differences are infinite, as in this model of
usual phantom there is a unique big rip.

Given that, as we have said before, the infinite singularities
have cut off the spacetime generating infinite causally
disconnected spacetimes, we can re-scale and redefine the time in
each of these spacetimes in some appropriate form, independently
in each of them. This way, the scale factor reaches its minimum
value in the zero of the so obtained new symmetrical time of
symmetry. The aforementioned scale factor can be written in a more
compact form as
\begin{equation}\label{f}
a(\tau)=a_{{\rm min}}\left({\rm
cos}\tau\right)^{-\frac{2}{3|\beta|}},
\end{equation}
with the new time $\tau$ covering the interval $(-\pi/2,\pi/2)$ in
every universe, reaching the initial and final big rips at the
extrema. Each of the universes in the multiverse is something as
though it were a faster-expanding de Sitter space defined along a
finite time interval.

If we assumed that all these universes are classically identical
and that our universe is in fact described by this model, we could
dare to claim that such universes are governed by the same
physical laws as ours, given that all of them would then be
exactly physically equivalent. Classically, the existence of life
in our universe might be justified as a byproduct of the anthropic
principle in its various formulations. If we think that life
exists because the initial conditions of our universe allow it to
occur, the physical equivalence of the various universes would
imply that, classical life existed such as we know it in all of
them. But if we considered the emergence of life as a process
somehow dependent on quantum effects, as it seems to be the case,
it would no longer be consistent to extrapolate ideas about such
existence based on a classical extension of the physical laws.

We could envisage a model where the expansion is not caused by a
phantom fluid, but by dark energy itself, i. e., $\beta>0$ in the
equation of state. In this case, we would also obtain a multiverse
scenario with the same characteristics among the universes, but
these would now be closed universes that would decelerate
from a big bang until its scale factor reached a finite maximum
value (given by Eq.(\ref{seis}) with $\beta>0$), from that value
onwards the universe would contract in size until finally it died
in a big crunch singularity (see Fig.~\ref{univ}); being
therefore unable to explain the current accelerated expansion of
our universe.

In view of the results obtained in this work, it is worth
mentioning that whereas the insertion of a negative cosmological
constant in a phantom energy model has the effect of repeating the
big rip singularity an infinite number of times, the analogous
consideration of this in a model with dark energy slows down the
accelerated expansion caused by this fluid, in such a way that it
would cause not just one but infinite big crunches. Hence in both
cases we obtain a classical multiverse scenario, in which the
universes are identical among them. This scenario could be altered
if we included the evolution of astronomical objects in this model
\cite{Yurov}.

As we said before, the models suggested in the present paper are
purely classical, therefore considering quantum effects would
probably smooth out the singularities \cite{Nojiri:2004ip}, in such a way that we would
no longer have an infinite set of isolated spacetimes, so implying
the loss of the multiverse scenario.

The appeal of the multiverse models lies on that it points toward
a less predominant position of what we call our universe in
nature. It could well be that, once again, we would have missed
the denomination of a physical system and, in a similar way to
terms such as atom or elementary particles were once wrongly used
to denote what it turned out to be essentially divisible systems,
we could well be now applying the term "universe" to what is
nothing but just a single part or product of it \cite{Pedro1986}.

\acknowledgements
The authors want to thank A.~Yurov for useful comments. This work was supported by DGICYT under Research Project No.~FIS2005-01180. P.~M.-M.~ and A.~R.-F.~ acknowledge CSIC and ESF for a I3P grant and MEC for a FPU grant, respectively.

\end{document}